\documentclass[preprint,showpacs,preprintnumbers,amsmath,xamssymb,pop]{revtex4-1}

\usepackage{bm} 
\usepackage{amssymb,amsmath,amsfonts,amstext,}

\def\bq{\begin{equation}}
\def\eq{\end{equation}}
\def\bqy{\begin{eqnarray}}
\def\eqy{\end{eqnarray}}

\def\calj{\mathcal{J}}

\def\de{\delta}
\def\na{\nabla}
\def\Om{\Omega}

\def\bfb{\mathbf{B}}
\def\bfe{\mathbf{E}}
\def\bfx{\mathbf{x}}
\def\bfv{\mathbf{v}}
\def\ua{\mathbf{a}}
\def\bfq{\mathbf{q}}

\def\hb{\hat{\mathbf{b}}}
\def\hv{\hat{\mathbf{v}}}
\def\hw{\hat{\mathbf{w}}}

\def\bbb{\bar{\bar{\mathbf b}}}

\def\ipd{\bar{\bar{I}}_{\perp}\cdot}
\def\dip{\cdot\bar{\bar{I}}_{\perp}}
\def\ip{\bar{\bar{I}}_{\perp}}

\def\cipd{\bar{\bar{\mathcal{I}}}_{\perp}\cdot}
\def\cdip{\cdot\bar{\bar{\mathcal{I}}}_{\perp}}

\def\x{\times}
\def\p{\partial}

\def\hp{\p_{\hv}}
\def\hpw{\p_{\hw}}
\def\pa{\p_{A}}
 
\def\pp{\p_{||}}
\def\ppr{\p_{\perp}}

\begin{document}

\title{Magnetic moment type of lifting  from particle dynamics to Vlasov-Maxwell dynamics}
\author{P. J. Morrison}
\email{morrison@physics.utexas.edu}
\affiliation{Department of Physics and Institute for Fusion Studies, University
  of Texas at Austin, Austin, TX 78712, USA}
\author{M.~Vittot}
\email{michel.vittot@cpt.univ-mrs.fr}
\affiliation{Centre de Physique Theorique\\  
Aix-Marseille Universite, CNRS, CPT\\
 UMR 7332, 13288 Marseille, France\\
and\\
Universite de Toulon, CNRS, CPT\\
 UMR 7332, 83957 La Garde, France}
\author{L. de Guillebon}
\email{loic.de-guillebon@cpt.univ-mrs.fr}
\affiliation{Centre de Physique Theorique\\  
Aix-Marseille Universite, CNRS, CPT\\
 UMR 7332, 13288 Marseille, France\\
and\\
Universite de Toulon, CNRS, CPT\\
 UMR 7332, 83957 La Garde, France}

\begin{abstract}

Techniques for coordinate changes that depend on both dependent and independent variables  are developed and applied to the Maxwell-Vlasov Hamiltonian theory.   Particle coordinate changes with a new velocity variable dependent  on the magnetic field,  with  spatial coordinates  unchanged,  are lifted to transform the  noncanonical Poisson bracket and, thus,  the field Hamiltonian structure of the Vlasov-Maxwell equation.  Several examples are given including  magnetic coordinates, where the velocity is decomposed into  components parallel and perpendicular to the local magnetic field, and the case of spherical velocity coordinates.  An example of the lifting procedure is performed to obtain a  simplified version of gyrokinetics, where the magnetic moment is used as a coordinate and the dynamics is reduced by elimination of the electric field energy in the Hamiltonian.   

\bigskip

\end{abstract}

\maketitle

%%%%%%%%%%%%%%%%%%%%%%%%%%%%%%%
\section{Introduction}
\label{sec:intro}

Perturbation theory in the context of Hamiltonian dynamics has proved to be unquestionably useful in many contexts, ranging from celestial mechanics (e.g. \cite{celestial}), to atomic physics (e.g.\ \cite{born}), to plasma physics (e.g.\ \cite{kruskal}).  The  superconvergent expansions of Kolmogorov-Arnold-Moser theorem (e.g.\ \cite{llave}) and the techniques of adiabatic invariance (e.g.\ \cite{henrard}) all are aspects of perturbation theory in the  Hamiltonian context. 
Although such techniques are well-developed and well-known for  finite-dimensional systems, this is not the case for  such perturbation theories for partial differential equations.  This is  particularly true for Hamiltonian systems with noncanonical Poisson brackets of the form of those given in \cite{morgreene,Morr81,morrison98} for plasma systems. A  main goal of the present paper is to provide tools for such perturbation theory using the Poisson bracket for  Vlasov-Maxwell equations \cite{Morr80, Wein81,Mars82, Morr81,morrison12}  in situations with a short time scale introduced by the presence of a strong magnetic field.

Derivations of gyrokinetic theories have proceeded directly from the Vlasov-Maxwell equations of motion as in the nonlinear development of \cite{Frieman},  they have been based on Hamiltonian particle orbit perturbation theory that is lifted up to the kinetic level as in the linear  development of \cite{littlejohn84}, or they have incorporated   both particle orbit and kinetic perturbations  to arrive at  a nonlinear theory \cite{mischenko}.  (See \cite{CaryBriz09, BrizHahm07} for review.)  None of these procedures parallels  that for  finite-dimensional Hamiltonian systems that has historically achieved such great success.  Consequently, none  of these theories obtain an infinite-dimensional Hamiltonian form as a consequence  of their method of derivation.  In fact, at present it is not known  if  nonlinear gyrokinetics  has Hamiltonian form,  the form possessed by  all of the important systems of plasma physics when dissipative terms are neglected. (For review of Hamiltonian structure and techniques see \cite{Morr81,morrison98,Morr05}). 

To effect an infinite-dimensional Hamiltonian gyrokinetic-like perturbation theory requires a sequence of coordinate changes that involves both the dependent (field) variables and their arguments, which are  independent variables  from the point of view of the Hamiltonian structure.    This complicates matters significantly and care must be taken  when performing transformations, most notably with the chain rule.  Because the Vlasov-Maxwell theory has fields of mixed type, the electromagnetic fields depending on a  space  variable and the distribution function depending  on a phase space variable, and because these fields are not a usual canonical set, the situation is further complicated.    In the present paper  the intricacies of this kind of transformation and associated chain rule are described, which enables the Hamiltonian perturbation theory.   The techniques are then  applied to obtain a  simplified version of gyrokinetics (guiding center kinetics), which considers the presence of a  conserved magnetic moment,  as a first step for more general  gyrokinetic reduction, e.g.\  by using the intrinsic coordinates developed in  \cite{IntrGyro},  that will be considered elsewhere. 

 The paper is organized as follows. In Sec.~\ref{sec:prelim} some preliminary material needed for the subsequent development is described.  This is followed by four sections  where several specific transformations are considered.  It is shown  how to lift these  coordinate transformations, which are  tailored to particle orbit dynamics,  up to the level of fields by detailing  how to transform the Vlasov-Maxwell Poisson bracket into the new coordinates.   Lifting in this context is a natural relative of that treated in Ref.~\cite{morrison12},  a companion paper that treats  the lifting of microscopic particle dynamics up to the field level.  Section \ref{sec:mag-coords} considers  magnetic coordinates, where the particle velocity coordinate is projected   parallel and perpendicular to a space-dependent dynamic magnetic field.  Because  the new particle coordinates depend on the  field, two chain rules must be considered:  the usual function chain rule for phase space coordinates,  where the field is assumed to be a given,  and the chain rule for functionals which is needed  for transforming the field theoretic Poisson bracket.   Next, in Sec.~\ref{sec:spherical}, spherical velocity coordinates are considered.  Here the velocity coordinates are chosen as the unit vector of the velocity (independent of the spatial coordinates) and a coordinate in one-to-one correspondence  with the norm of the velocity. This transformation is a  step closer to that needed for  gyrokinetics where one introduces the  magnetic moment,  and it introduces a new feature in that  the Jacobian determinant of the transformation is no longer unity.  Thus,  it provides a simple example  for  explaining how functional derivatives change when Jacobians change.  In Sec.~\ref{sec:local}, we turn to a more complete  case where the change of coordinates depends only on the local value of the magnetic field,  as a precursor to Sec.~\ref{sec:nonlocal} that considers  the physically important  situation, where the change of coordinates involves spatial derivatives of the magnetic field to arbitrary order, i.e., as given by Eq.~(\ref{gentrans}) below.   With the techniques of the  previous four sections in hand, in Sec.~\ref{sec:example} we treat an example  where the reduced coordinate is indeed the magnetic moment and explicitly transform the Hamiltonian form of the Vlasov-Maxwell equation into  the  new coordinates.   Finally, in  Sec.~\ref{sec:conclusion}, we conclude.  

 \section{Preliminary material}
 \label{sec:prelim}

From a general point of view, a main  purpose  of this paper is to transform the Vlasov-Maxwell Hamiltonian structure when the phase space variables $(\bfq,\bfv)$ are changed to the following new coordinates that depend on the  magnetic field and all of its derivatives:
\bq
\bar \bfq = \bfq
\,,\qquad 
\bar \bfv=  \bar\bfv(\bfq, \bfv; \bfb,\nabla\bfb,\dots)\,.
\label{gentrans}
\eq

 For the noncanonical Hamiltonian structure of Vlasov-Maxwell dynamics, the observables are the set of all functionals of the magnetic field $\bfb (\bfq)$,  the electric field $\bfe (\bfq)$, and  the phase space  density $f (\bfq,\bfv)$, where the time variable has been suppressed. The Poisson bracket is \cite{Morr80, Wein81, Mars82, Morr81}:
\bqy
	\{{F},{G}\}&=&\int\!\!d^3q d^3v\, f~ [ F_f,{G}_f] 
	\label{BrackMV}
	\\
	&+&
	e\int\!\!d^3q d^3v~  f 
	\left(  G_{\bfe}\cdot\p_{\bfv}  F_f  
	-  F_{\bfe}\cdot\p_{\bfv}  G_f\right)
	\notag\\
	&  +& 
	\int\!\!d^3 q ~ \big( F_{\bfe}\cdot \nabla\times   G_{\bfb}
	-  G_{\bfe}\cdot \nabla\times   F_{\bfb}\big)
	\,,
	\notag
\eqy
where subscripts are used for  functional derivatives,     ${F}_f:=\de  F/\de f$,    ${F}_{\bfe}:=\de  F/\de \bfe$,  etc.,  and the particle bracket is $[f,g] =\nabla f\cdot \p_{\bfv}   g -\nabla g\cdot \p_{\bfv}  f + e\bfb\cdot \p_{\bfv} f \times \p_{\bfv} g$, with $\nabla f=\p f/\p \bfq$ and $\p_{\bfv} g=\p g/\p \bfv$. For the sake of simplicity physical constants  have been scaled away as usual, but  a dimensionless charge variable  $e$ that indicates the coupling term  has been retained (see \cite{morrison12} for a dimensional form of this bracket). 
The variable $e$ becomes the charge ratios when  (\ref{BrackMV}) is generalized by summing over  multiple species.

The Hamiltonian functional is 
\bq
	 H[\bfe,\bfb,f] = \tfrac1{2}  \int\!\!d^3q d^3v\, \|\bfv\|^2   f + \tfrac1{2}\int\!\!d^3q\, \left(\|\bfe\|^2+|\bfb\|^2\right)
	 \,,
	\label{ham}
\eq
which is the sum of the kinetic energy of the plasma and  the  energy of the electromagnetic field.  The  relativistic model is obtained by replacing $\|\bfv\|^2$ in the kinetic energy term with $\sqrt{1 + \|\bfv\|^2}$,  where in the latter case $\bfv$ is the scaled relativistic momentum.   The coupling between the plasma and electromagnetic field is included in the noncanonical Poisson bracket  (\ref{BrackMV}). The Hamiltonian (\ref{ham}) together with the Poisson bracket generates the motion through Hamilton's equations  expressed as 
\bqy
	\dot F &=&
	\{{F}, H\}
	\,,
	\notag
\eqy
for any observable $F$. In particular, if $F$ denotes the field variables the bracket  induces Maxwell-Vlasov equations as follows:
\bqy
	\dot \bfb &=&\{\bfb,H\}= -\na\x\bfe
	\,,\notag \\
	\dot \bfe &=&\{\bfe,H\}=\na\x\bfb - e\int\!\!d^3v~f\bfv
	\,,\notag \\
	\dot f &=&\{f,H\}=-\bfv\cdot\na f 
	- e~(\bfe+\bfv\x\bfb)\cdot\p_{\bfv} f
	\,.
	\notag
\eqy

As noted in Sec.~\ref{sec:intro}, in order to  transform the Hamiltonian structure to facilitate the separation or removal of fast time scales (as in oscillating-center, guiding-center, and gyrokinetic theories) care must be taken because such a change of coordinates involves both the dependent and independent variables, i.e.,  the spatial observation points of the  field.  A simple case of this is treated in the next section. 
%
%If the particle orbit theory was one in terms of canonical variables, then the formalism of \cite{TheoryLift10} would be available.  However, when  a perturbation expansion has been employed in order e.g.\ to facilitate separation or removal of fast time scales (as in oscillating-center, guiding-center, gyrokinetic theories), then the resulting equations of motion are often noncanonically Hamiltonian, i.e.\  possess a noncanonical Poisson bracket \cite{Litt82}. When this is the case difficulties can occur when attempting to find a Hamiltonian theory  because general phase space transformations can mix the spatial observation points of the electromagnetic field variables. It is the reason why we consider here two examples of {\it homogeneous lifts}, where this difficulty does not occur because only the particle velocity coordinates are changed. We let for a future work the general case where the particle position is changed as well. 
%
 
 %%%%%%%%%%%%%%%%%%%%%%%%%%%%
\section{Lifting with magnetic coordinates}
 \label{sec:mag-coords}

As a first case of  lifting,  consider velocity coordinates  based on a decomposition of the velocity using the magnetic field. This transformation of the spatial coordinate is unchanged, but the velocity  $\bfv$ is transformed as follows:
\bq
	\bfv
	=  \bfv(\bar \bfv; \bfb)
	= \bfv(v_{||},  \bfv_{\perp}; \bfb)
	=\hb v_{||} + \bfv_{\perp}
	\,,
	\notag
\eq
where $\hb=\bfb/\|\bfb\|$ is the unit vector the direction of the magnetic field, 
\bq
	v_{||}=\hb\cdot\bfv 
	\notag
\eq
is the (scalar) component of the velocity parallel to the magnetic field, and 
\bq
	\bfv_{\perp} = \bfv-\hb\hb\cdot\bfv =\ipd\bfv
	\notag
\eq
is the (vectorial) component of the velocity perpendicular to the magnetic field, with 
\bq
	\bar{\bar{I}}_\perp:= \bar{\bar{I}}- \hb\hb
	\label{PerpProjMagn}
\eq
being the orthogonal projector onto the plane perpendicular to the magnetic field.

There are two chain rules to consider:  that for functions, considered next, and that for functionals, such as the energy expression of (\ref{ham}), which will follow.

 \subsection{Function chain rule}

The transformation of the field Poisson bracket of (\ref{BrackMV}) requires the transformation of the particle bracket, 
\bq
	[f,g]=\frac{\p f}{\p \mathbf{q}}\cdot\frac{\p g}{\p \bfv}
	- 
	\frac{\p g}{\p \mathbf{q}}\cdot\frac{\p f}{\p \bfv}
	+ e \bfb\cdot\left(\frac{\p f}{\p \bfv}
	\times\frac{\p g}{\p \bfv}\right)\,, 
	\label{ptlbkt}
\eq
into the new coordinates,  
$(\bfq,\bfv)\rightarrow(\bfq,v_{||},\bfv_{\perp})$. 
The following abbreviations are convenient:
\bq
	\nabla
	:=\frac{\p f}{\p \mathbf{q}}
	\,,\quad \p_i=\frac{\p f}{\p{q}_i}\,,
	\quad
	\p_{||}= \frac{\p f}{\p v_{||}}\,,
	\quad
	\p_{\perp}= \frac{\p f}{\p \bfv_{\perp}}
	\,.
	\notag
\eq
Note the last operator acts only in the plane perpendicular to $\bfb$, which implies the following properties: 
\bq
	\ppr \bar{f}\dip=\ppr\bar{f} \qquad 
	{\rm and}\qquad \hb   \cdot\ppr\bar{f}=0\,.
	\notag
\eq

 Total variations of $f(\bfq,\bfv)=\bar{f}(\bfq,v_{||},\bfv_{\perp})$ are given by  
\bq
	\de f= \frac{\p f}{\p \bfq}\cdot\de\bfq 
    + \frac{\p f}{\p   \bfv}\cdot\de\bfv
    = 
    \nabla \bar{f}\cdot\de\bfq+ \pp\bar{f}.\de v_{||} 
    + \ppr \bar{f}\cdot\de\bfv_{\perp}\,, 
	\label{df}
\eq
while variations of the initial and final coordinates are related by
\bqy
    \de v_{||}
    &=& \hb \cdot \de\bfv
    + (\de\bfq\cdot\nabla \hb)\cdot \bfv
    \,,
	\nonumber\\
    \de\bfv_{\perp}&=& \ipd\de\bfv-\de\ipd \bfv
    \nonumber\\
	&=&\ipd\de\bfv
	-  (\de\bfq\cdot\nabla \hb)~(\hb\cdot\bfv) \nonumber\\
	&&\hskip10ex-   \hb ~ 
	(\de\bfq\cdot\nabla\hb) \cdot\bfv
    \,.
    \label{delts}
\eqy
For the function chain rule the field $\bfb$ is assumed to be a fixed function with  the coordinates $(\mathbf{q}, \bfv)$ changing.

 Inserting (\ref{delts}) into (\ref{df}) implies the chain rule relations 
\bqy
	\frac{\p f}{\p \bfv}
	&=&\hb~  \pp \bar{f} 
	+ \ppr\bar{f}\dip
	=\hb~  \pp \bar{f} + \ppr\bar{f} 
	\,,
	\label{dv}
	\\
	\frac{\p f}{\p q_i}&=&\p_i\bar{f} 
	+ (\bfv\cdot\p_i\hb)~ 	\pp\bar{f}
	-(\hb\cdot\bfv)~\ppr\bar{f} \cdot\p_i \hb
	\,, 
	\label{dq}
\eqy
and using (\ref{dv}) and (\ref{dq}) in (\ref{ptlbkt}) gives the particle bracket in the magnetic coordinates 
\bqy
	[\bar f,\bar g]
	&=&\hb\cdot
	\left(\nabla \bar f~  \pp \bar g -\nabla \bar g~  \pp \bar f\right)
	\label{transbkt}
	\\
	&+& \left(\nabla \bar f\cdot \ppr \bar g 
	-\nabla \bar g\cdot  \ppr \bar f\right)
	\nonumber\\
	&+& \ua\cdot 
	\left(\ppr \bar g~  \pp \bar f 
	-\ppr \bar f~  \pp \bar g\right) 
	\nonumber\\
	&+&\ppr \bar f\cdot \bbb\cdot\ppr \bar g
	+e\bfb\cdot \left(\ppr \bar f \times\ppr \bar g\right)
	\,,
	\nonumber
\eqy
with 
\bq
    a_i=\bfv\cdot \p_i\hb+ (\hb\cdot\bfv)~ \hb\cdot\p \hb_i\quad
    {\rm and}\quad
    {\bar{\bar{b}}}_{ij}
    = (\hb\cdot\bfv)
    \left( \p_i \hb_j-  \p_j \hb_i\right)\,.
    \notag
\eq
In all these relations, recall  that $\ppr\bar{f}=\ppr \bar{f}\dip$.  This is important because, for instance, the component of $\na \bar g$, $\ua$ or $\bbb$ parallel to $\hb$ are non-zero, but vanish when contracted with $\ppr \bar f$.

 \subsection{Jacobian}

 In general care  must be taken with the Jacobian determinant $\calj$ when defining functional derivatives, but here the Jacobian is unity 
\bq
	\calj:=\frac{\p(\bfq,v_{||},\bfv_{\perp})}{\p(\bfq,\bfv)}= 1
    \,.
    \notag
\eq
This follows because rotations have unit Jacobians and at any time there exists a rotation to a cartesian coordinate system with one of the $\bfv$ axes aligned with $\hb$. Thus
\bq
	dz:= d^3qd^3v
    =d^3qdv_{||}d^2v_{\perp}
    =: dq dv\,.
    \notag
\eq
  
 Because the volume integral is ultimately independent of how it is calculated,  $dz$ can be assumed to be independent of $\bfb$, e.g.\ when calculating functional derivatives with respect to $\bfb$, the topic considered next.

 \subsection{Functional chain rule}

 For the functional chain rule, the transformation of the fields must be made definite,  Here, 
\bqy
	\bfe(\bfq)&=&\bar\bfe(\bfq)
    \,,\qquad
    \bfb(\bfq)=\bar\bfb(\bfq)
    \,,
    \notag
    \\
    f(\bfq,\bfv)
    &=&\bar{f}(\bfq,v_{||},\bfv_{\perp})
    = \bar{f}(\bfq,\hb\cdot\bfv,\ipd\bfv)
    \nonumber\\
    &=& f(\bfq,\hb v_{||}
    +\bfv_{\perp})
    \,,
    \notag
\eqy
where now the coordinates $(\mathbf{q}, \bfv)$ are fixed and the field $\hb$ varies.

Variation of a transformed functional,  $F[f,\bfb,\bfe]=\bar{F}[\bar{f},\bar\bfb,\bar\bfe]$, gives 
\bqy
	\de F
	&=&\int\!\! dz~  F_f\de f
	+\int\!\! dq ~ \left(F_{\bfb}\cdot\de\bfb 
	+{F}_{\bfe}\cdot\de{\bfe} \right)
	\nonumber\\
	&=&\int\!\! dz~  \bar{F}_{\bar f}\de \bar{f}
	+\int\!\! dq ~ 
	\left(\bar{F}_{\bar{\bfb}}\cdot\de\bar{\bfb} 
	+ \bar{F}_{\bar\bfe}\cdot\de\bar{\bfe}\right)
	\,.
	\label{deF}
\eqy

 With the variations of the initial and final fields related by
\bq
	\de\bfe
	=\de\bar{\bfe}
	\,,\quad 
	\de\bfb=\de\bar{\bfb}
	\,,\quad {\rm and}\quad 
	\de f
	=\de\bar{f} 
	+ \pp\bar{f} ~(\bfv\cdot\de\hb) 
	+\ppr\bar{f}\cdot\de \ipd\bfv
	\,,
	\label{defn}
\eq
expressions relating functional derivatives of new and old variables  can be obtained. Using
\bq
	\de \ip 
	= -\frac1{\|\bfb\|}\left(\hb~  \ipd\de \bfb 
	+ \ipd\de \bfb~  \hb\right)
	\,,
	\notag
\eq
and after some work the last equation of (\ref{defn}) becomes
\bq
	\de f
	=\de\bar{f} 
	+\frac{(\bfv_{\perp}\cdot\de\bfb)}{\|\bfb\|}~   \pp\bar{f}  
	- \frac{v_{||}}{\|\bfb\|}~ \de \bfb 
	\cdot\ppr\bar{f} \,.
	\notag
\eq
Inserting this and the other two equations of (\ref{defn}) into (\ref{deF}),  and then equating coefficients, gives the functional chain rule relations 
\bqy
	\frac{\de F}{\de f}
	&=&\frac{\de \bar{F}}{\de \bar{f}}
	\,,\quad   
	\frac{\de F}{\de \bfe}
	=\frac{\de \bar{F}}{\de \bar{\bfe}}
	\,,
	\notag
	\\
	\frac{\de F}{\de \bfb}
	&=&\frac{\de \bar{F}}{\de \bar{\bfb}} 
	+ \frac1{\|\bfb\|} \int\!\! dv~ 
	\frac{\de F}{\de \bar f}~ \p_{\bfv}^* \bar{f}
	\,,
	\label{chainp2}
\eqy
where  
\bq
	\p_{\bfv}^*:=\bfv_{\perp} \pp  - v_{||} \ppr 
	\,.
	\label{pvs}
\eq

 Finally, the Maxwell-Vlasov bracket expressed in these magnetic coordinates is 
\bqy
	\{{F},{G}\}
	&=&\int\!\!dz~ f~  [ F_f,{G}_f]  
	\notag
	\\
	&+& e\int\!\!dz~ f \left(  G_{\bfe}\cdot\p_{\bfv}  F_f  
	-  F_{\bfe}\cdot\p_{\bfv}  G_f\right)
	\label{tdbkt}
	\\
	&{+}& \int\!\!d^3q ~ 
	\left( F_{\bfe}\cdot \nabla\times  \left[ G_{\bfb} 
	+ \frac1{\|\bfb\|} \int\!\! dv~ 
	G_f ~ \p_{\bfv}^* {f}
	\right]\right.
	\nonumber
	\\
	&&\left.-  G_{\bfe}\cdot \nabla\times  \left[ F_{\bfb} 
	+ \frac1{\|\bfb\|} \int\!\! dv~ F_f ~ \p_{\bfv}^*{f}
	\right]\right)\,,
	\nonumber
\eqy
where the `bars' have been dropped,    $[\, ,\, ]$  means the  bracket of (\ref{ptlbkt}) rewritten in the new coordinates as (\ref{transbkt}), and $\p_{\bfv}=\hb~  \pp   + \ppr $  is a shorthand  as in  (\ref{dv}).  Note, $\p_{\bfv}^* v^2=0$.

 %%%%%%%%%%%%%%%%%%%%%%%%%%%%%%%
\section{Lifting spherical velocity coordinates $\bfv= V\hv$}
 	\label{sec:spherical}

 Now turn to the new coordinates considered for  intrinsic gyrokinetics (used in \cite{IntrGyro}), which changes only one of the velocity coordinates to get the magnetic moment. The two other velocity coordinates are usually chosen as the unit vector of the velocity. So, a preliminary change of coordinates consists in adopting spherical coordinates for the velocity space: $\bfv= V\hv$ where $V:=\left\Vert \bfv \right\Vert\in\mathbb R_+$ is the norm of the velocity and $\hv:={\bfv}/{\left\Vert {\bfv} \right\Vert}\in S^2$ is the unit vector of the velocity. This transformation is considered  in this section, but later    the change $V\rightarrow \mu$ will be  considered. 

 The transformation $\bfv\leftrightarrow (\hv,V)$ is clearly invertible.  For the chain rule  the following are needed:
\bq
	\de V= \hv\cdot\de\bfv
    \quad {\rm and}\quad
    \de\hv= \cipd \frac{\de\bfv}{V}\,,
    \notag
\eq
 where
\bq
    \bar{\bar{\mathcal{I}}}_{\perp} = \bar{\bar{I}}-\hv\hv
    \notag
\eq
is  the orthogonal projector onto the plane perpendicular to the velocity. Note $ \bar{\bar{\mathcal{I}}}_{\perp} $  is different from the magnetic projector $\bar{\bar{I}}_\perp$ of (\ref{PerpProjMagn}) used in Sec.~\ref{sec:mag-coords}. 

 As in Sec.~\ref{sec:mag-coords}  the above are used to calculate the function chain rule, giving 
\bqy
    \frac{\p f}{\p \bfv
    }&=&\frac1{V}\frac{\p \bar{f}}{\p \hv}   \cdip   
    +  \frac{\p \bar{f}}{\p V}~ \hv 
    \,,
    \label{ddv}
    \\
    \nabla f 
    &=&  \frac{\p f}{\p \bfq}
    =  \frac{\p \bar{f}}{\p \bfq}
    = \nabla\bar f
    \,.
    \label{Vchain}
\eqy
 
 Inserting (\ref{ddv}) and (\ref{Vchain}) into (\ref{ptlbkt}) and, after some manipulations,   the particle bracket expressed in spherical coordinates  is obtained 
\bqy
	[f,g]
	&=&\frac1{V}
	\left(\nabla f\cdip \cdot \hp g
	- \nabla g\cdip \cdot \hp f \right)
    \notag\\
    &+& \hv\cdot\left(\nabla f ~\p_V g- \nabla  g ~\p_Vf \right)
    \nonumber
    \\
    &+&
    \frac{e\bfb}{V^2}\cdot
    \left(\hp f\cdip\right)
    \times
    \left(\hp g\cdip\right)
    \nonumber
    \\
    &+&
    \frac{e\bfb\times \hv}{V}\cdot  
    \left(\p_V f~\hp g - \hp f ~\p_V g \right)  
    \,, 
    \label{nuinner}
\eqy
where, for convenience,  the `bars' have been dropped and the abbreviations 
\bq
    \frac{\p f}{\p \hv}=:\hp f
	\quad{\rm and}\quad 
	\frac{\p f}{\p V}=:\p_Vf
	\,,
    \notag
\eq
have been employed.
  
 Turning to the functional chain rule,   notice that the change of coordinates does not depend on the fields, but the Jacobian for this special case is not unity
\bq
    dz
    = V^2 dVd\Omega d^3q 
    =\calj dVd\Om d^3q=:\calj d\eta d^3q
    =: \calj dw
    \,,
    \label{jac}
\eq
because the integration measures are changed from $d^3v$ and $dz$ to $d\eta$ and $dw$, which are defined by relation (\ref{jac}).
 
Thus, as above, 
\bqy
    \de F&=& \int\!\! dz ~ F_f \de f
    +\int\!\! d^3q  ~ \left(F_{\bfb}\cdot\de\bfb 
    +{F}_{\bfe}\cdot\de{\bfe} \right)
	\label{first} \\
    &=& \int\!\!  dw   ~  \bar{F}_{\bar f}\de \bar{f} 
    + \int\!\! d^3q   ~ 
    \left( \bar{F}_{\bar{\bfb}}\cdot\de\bar{\bfb} 
    + \bar{F}_{\bar\bfe}\cdot\de\bar{\bfe}\right)
    \,.
    \notag
\eqy
Inserting (\ref{jac}) into (\ref{first}) gives
\bq
    F_f=\calj ^{-1}\bar{F}_{\bar f}
    \,,\quad 
    F_{\bfb} = \bar{F}_{\bar{\bfb}}
    \,,\quad{\rm and}\quad 
    {F}_{\bfe}=\bar{F}_{\bar\bfe}
    \,.
    \label{jfuncd}
\eq
 Note,  in (\ref{jfuncd})  the new functional derivative is defined with respect to the {\it bare} measure $dw$.  
 
 So, the first term of the Maxwell-Vlasov bracket transforms as 
\bq
    \{F,G\}_1
    :=\int\!\! dz~ f\left[F_f,G_f\right]
    = \int\!\! dw ~ 
    \calj \bar{f}
    \left[
    \calj^{-1}\bar{F}_{\bar{f}}
    ,
    \calj^{-1}\bar{G}_{\bar{f}}
    \right]
    =\{\bar{F},\bar{G}\}_1\,,
    \label{jvp}
\eq
with  the bracket of  the second equality  above  given by (\ref{nuinner}). 
 
 The basic identity for this bracket with Jacobians, which replaces the usual `$f$-$g$-$h$' identity for canonical brackets $\int \!dz \, f[g,h]=-\int \!dz\,  g[f,h]$,  is the following:
\bq
    \int\!\! dw~  \calj f\left[\calj^{-1}g,h \right]
    = -  \int\!\! dw ~ g\left[f ,h \right]
    \,.
    \label{jfgh}
\eq
 In terms of the bare measure
\bq
    \frac{\de f(w)}{\de f(w')}=\de(w-w')\,.
    \label{barede}
\eq
 The bracket of (\ref{jvp}) with (\ref{jfgh}) and (\ref{barede}) produces the correct equations of motion for the Vlasov-Poisson system. 
 
 Now consider the coupling terms of the bracket
\bqy
	\{F,G\}_2
    &:=&
    e\int\!\!dz~ f \left(  G_{\bfe}\cdot\p_{\bfv}  F_f  
    -  F_{\bfe}\cdot\p_{\bfv}  G_f\right)
    \notag\\
    &=&
    e\int\!\!dw~ \calj \bar{f} 
    \left(  
    G_{\bar\bfe}\cdot\p_{\bfv}  
    \calj^{-1}\bar{F}_{\bar f}  
    -  F_{\bar\bfe}\cdot\p_{\bfv}  
    \calj^{-1}\bar{G}_{\bar f}
    \right)
    \,,
	\notag
\eqy
where $\p_{\bf v}$ is a shorthand for the expression of (\ref{ddv}).  When generating  Maxwell's equations, the Hamiltonian gives
\bq
	\bar{H}_{\bar f}
	= \calj \|\bfv\|^2/2\,, 
	\notag
\eq
which  gives the correct expression for the current  density $\mathbf J=\int\! d\eta \, \calj f \bfv$. 

 Finally,  the pure field terms of the Maxwell-Vlasov bracket are unchanged and, thus,  the Maxwell-Vlasov bracket in these spherical coordinates becomes 
\bqy
	\{{F},{G}\}
	&=&
	\int\!\! dw ~ \calj f
	\left[\calj^{-1}F_f,\calj^{-1}G_f\right]
	\notag
	\\
	&+&
	e\int\!\!dw~ \calj f 
	\left(
	  G_{\bfe}\cdot\p_{\bfv}  \calj^{-1} F_f  
	-  F_{\bfe}\cdot\p_{\bfv}  \calj^{-1} G_f 
	\right)
	\notag\\
	&+&
	\int\!\!d^3 q ~ \big( F_{\bfe}\cdot \nabla\times   G_{\bfb}
	-  G_{\bfe}\cdot \nabla\times   F_{\bfb}\big)
	\,,
	\notag
\eqy
where the `bars' have been dropped, and $[\, ,\, ]$  means the  bracket of (\ref{ptlbkt}) rewritten in the new coordinates as (\ref{nuinner}).

%%%%%%%%%%%%%%%%%%%%%%%%%%%%%%%
	\section{Lifting with  local dependence on $\bfb$}
	\label{sec:local}

 To include the magnetic moment in the coordinates, the next step is to investigate the coordinate transformation $V\leftrightarrow A$, where $A$ is a coordinate in one-to-one correspondence with  the  coordinate $V$ of Sec.~\ref{sec:spherical}, but  in this section  it is assumed to have local dependence on the  magnetic field, i.e, it depends on  $\bfb$  but not its derivatives.   Explicitly, the tranformation is 
$(\bfq, V,\hv)\leftrightarrow(\bar\bfq, A,\hw)$ where
\bq
	\bfq =\bar \bfq
	\,, \quad  
	\hv	=\hw
	\,,  \quad{\rm and} \quad  
	V=V(A,\hw,\bfb)\,.
	\notag
\eq
Clearly, invertibility requires $V_A:=\p V/\p A\neq 0$.  Since the first two equations above are identities,  eventually $ \hv$ will be used for $\hw$ and $\bfq$ for $\bar\bfq$.

 The Jacobian for this transformation is now
\bqy
	dz
	&=& V^2dV d\Om d^3q
	= V^2V_A dA d\Om d^3q
	\notag
	\\
	&=&\calj dA d\Om d^3q
	=: \calj d\eta d^3q
	=:\calj dw
	\,,
	\notag
\eqy
which define the Jacobian $\calj$ and the integration measures $d\eta$ and $dw$. Note that these  are not the same as those of Sec.~\ref{sec:spherical}, even though the same symbols are used. Furthermore, $\calj$ now depends on $\bfb$ and, hence, $\bfq$.  Also, $d\Om$ contains a portion of the Jacobian from cartesian coordinates, but one that is independent of $\bfq$.

Now the chain rule is  effected  on functions analogous to (\ref{dv})-(\ref{dq}) and (\ref{ddv})-(\ref{Vchain}) and on functionals analogous to (\ref{chainp2}) and (\ref{jfuncd}).  Varying $f(\bfq, V,\hv)=\bar{f}(\bar\bfq,A,\hw)$ in the label (coordinates) dependence, and then equating as above, gives
\bqy
	\frac{\p  f}{\p \bfq}
    &=&\frac{\p  \bar f}{\p\bar  \bfq}
    -\frac{V_{B_i}}{V_A}  \frac{\p  B_i}{\p \bfq}~ 
    \frac{\p\bar f}{\p A}
    \,,
    \label{dfdq}\\
    \frac{\p  f}{\p V}
    &=&\frac1{V_A}\frac{\p\bar f}{\p A}
    \,,
    \label{PV}
    \\
    \frac{\p  f}{\p \hv}&=&\frac{\p  \bar f}{\p \hw} 
    -\frac{1}{V_A} \frac{\p V}{\p \hv}~ 
    \frac{\p \bar f}{\p A}
    \,.
    \label{phv}
\eqy
Inserting (\ref{PV}) and (\ref{phv}) into (\ref{ddv}) gives the chain rule on functions 
\bqy
	D_* \bar{f}
	&=&\frac{\p  f}{\p \bfv}
	= \frac1{V} \left( \frac{\p  \bar f}{\p \hw}   
	- \frac{1}{V_A}\frac{\p \bar f}{\p A}~
	\frac{\p V}{\p \hw}  \right) \cdip 
    +  \frac{\hw}{V_A}\frac{\p\bar f}{\p A} \notag\\
    &=& \frac1{V} \hpw\bar f \cdip 
    +  \frac{ \pa\bar f}{V_A}~\hw
    -  \frac{\pa\bar f}{VV_A}  ~  \hpw V\cdip 
    \,,
	\label{newds}   
\eqy
while (\ref{dfdq}) gives 
\bq
	\nabla_* \bar{f}
	=\frac{\p    f}{\p \bfq}
	=\frac{\p  \bar f}{\p \bar \bfq} 
	- \frac{V_{B_i}}{V_A}  \frac{\p  B_i}{\p\bar \bfq}~
	\frac{\p\bar f}{\p A}
 	= \bar \nabla \bar f 
 	- \frac{V_{B_i}}{V_A} ~ \bar\nabla B_i ~ \pa\bar f
 	\,.
	\label{nabds}
\eq
Then,  inserting (\ref{newds}) and (\ref{nabds}) into (\ref{nuinner}) gives the following complicated expression for the particle bracket $[\ ,\ ]$ in the new coordinates:
\bqy
	[\bar f,\bar g]
	&=&
	\nabla_*  \bar f \cdot D_*  \bar g 
	-  \nabla_*  \bar g \cdot D_*  \bar f
	+  e\bfb\cdot\left( D_*  \bar f\times  D_*  \bar g\right)
	\notag
	\\
	&=&
	\frac1{V}\left(\bar\nabla\bar f\cdip \cdot \hpw \bar g
	- \bar\nabla \bar g\cdip \cdot \hpw \bar f \right)
	\notag
	\\
 	&+& \frac{\hw}{V_A}\cdot
 	\left(\bar\nabla \bar f~  \pa\bar g  
 	- \bar\nabla  \bar g~  \pa\bar f   \right)
	\notag
	\\
 	&+&\frac{\hpw V  \cdip}{VV_A} \cdot 
 	\left(\bar\nabla\bar g ~   \pa\bar f  
 	- \bar\nabla \bar f ~  \pa\bar g \right) 
 	\notag
 	\\
 	&+& \frac{V_{B_i}}{V V_A} ~  \bar\nabla B_i \cdip \cdot
 	\left( \hpw \bar f ~\pa \bar g
 	- \hpw \bar g ~\pa \bar f\right)
 	\nonumber
 	\\
 	&+&\frac{e\bfb}{V^2}\cdot
 	\left(\hpw \bar f \cdip\right)
 	\times
 	\left(\hpw \bar g\cdip\right)
 	\nonumber
 	\\
 	&+&
  	\frac{e\bfb\times \hw}{VV_A}
  	\cdot
  	\left( \hpw \bar g~  \pa\bar f 
  	- \hpw \bar f~  \pa\bar g
 	\right)   
  	\nonumber
 	\\
  	&-&\frac{e\bfb}{V^2V_A}
  	\times
  	\left(\bar{\bar{\mathcal{I}}}_{\perp}
  	\cdot \hpw V\right) 
  	\notag
  	\\
  	&&\hskip7ex
  	\cdot~
  	\bar{\bar{\mathcal{I}}}_{\perp}
  	\cdot 
  	\left( \hpw \bar g ~ \pa \bar f
  	-  \hpw \bar f ~ \pa \bar g \right)
  	\,.
 	\label{locinner}
\eqy

 Now consider the functional chain rule as above,
\bqy
 	\de F&=& \int\!\! dz ~ F_f \de f
 	+\int\!\! d^3q  ~ \left( F_{\bfb}\cdot\de\bfb 
 	+{F}_{\bfe}\cdot\de{\bfe} \right)
	\notag \\
 	&=& \int\!\!  dw   ~  \bar{F}_{\bar f}\de \bar{f} 
 	+ \int\!\! d^3q   ~ 
 	\left( \bar{F}_{\bar{\bfb}}\cdot\de\bar{\bfb} 
 	+ \bar{F}_{\bar\bfe}\cdot\de\bar{\bfe}\right)
 	\,,
 	\label{ddfda}
\eqy
Functionally varying $f(\bfq, V,\hv)=\bar{f}(\bar\bfq, A,\hw)$ gives
\bq
	\de f
	=\de\bar{f} 
	+ \frac{\p \bar f}{\p A}~ \frac{\p A}{\p \bfb} \cdot 
	\de \bfb \,,
	\label{ddfa}
\eq
while  $ \de \bfb =\de \bar{\bfb}$ and $\de \bfe=\de \bar{\bfe} $.  Whence, upon substitution of (\ref{ddfa}) into (\ref{ddfda}), the  chain rule on functionals  is obtained, 
\bqy
	\frac{\de F}{\de f}
	&=&
	\frac1{\calj }\frac{\de \bar{F}}{\de \bar{f}}
	\,, 
	\notag
	\\
	\frac{\de F}{\de \bfe}
	&=&
	\frac{\de \bar{F}}{\de \bar{\bfe}}
	\,,
	\notag
	\\
	\frac{\de F}{\de \bfb}
	&=&
	\frac{\de \bar{F}}{\de \bar{\bfb}} 
	-   \int\!\! d\eta~     
	\frac{\p A}{\p \bfb} ~  \frac{\p \bar f}{\p A}
	\frac{\de \bar F}{\de \bar f}  
	\,,
 	\label{chainp3}
\eqy
where the  last expression of (\ref{chainp3}) can be written in a more convenient way as
\bq
	\frac{\de F}{\de \bfb}
	=\frac{\de \bar{F}}{\de \bar{\bfb}}  
	+ \int\!\! d\eta~   
	\frac{V_{\bar \bfb}}{V_A}~  
	\frac{\p \bar f}{\p A}\frac{\de \bar F}{\de \bar f}  
	\,.
	\notag
\eq
This follows from 
\bq
	\frac{\p A}{ \p \bfb} 
	= -\frac{V_{\bar \bfb}}{V_A}\,,
	\notag
\eq
which comes about  because the change in $A$ induced by a change in $\bfb$ at fixed $V$ and $\hw$,  satisfies  
$0= \de V=V_A \de A + V_{\bar B_i}\de \bar B_i$. 

 Finally, the Maxwell-Vlasov bracket in the coordinates $(\bfq, A,\hv)$ is given by 
\begin{align}
 	&\{F,G\}
 	=  \int\!\! d\eta d^3q ~ \calj  {f}
 	\left[\calj^{-1}{F}_{{f}},\calj^{-1}{G}_{{f}}\right]
 	\label{BrackMVCoordA}
 	\\
  	&+    e\int\!\!  d\eta d^3q~ \calj {f} 
  	\left(  G_{\bfe}\cdot D_*  \calj^{-1}{F}_{ f}  
  	-  F_{\bfe}\cdot D_*  \calj^{-1}{G}_{ f}\right)
 	\nonumber
 	\\
	&+ \int\!\!d^3q ~ 
	\left( F_{\bfe}\cdot \nabla\times 
	\left[  G_{\bfb} 
	+  \int\!\! d\eta~   \frac{V_{ \bfb}}{V_A}~  
	\frac{\p f}{\p A}\frac{\de G}{\de f}
	\right]
	\right.
	\notag
	\\
	&\hskip5ex
	\left.
	-  G_{\bfe}\cdot \nabla\times  
	\left[ F_{\bfb}  
	+ \int\!\! d\eta~     \frac{V_{ \bfb}}{V_A}~  
	\frac{\p f}{\p A}\frac{\de F}{\de f}
	\right]
 	\right)\,,
	\nonumber
\end{align}
where the particle bracket $[\ ,\ ]$ is given by (\ref{locinner}),  $D_*$ is the operator defined by (\ref{newds}), and  the bars have been dropped.

%%%%%%%%%%%%%%%%%%%%%%%%%%%%%%%
	\section{Lifting with  nonlocal dependence on $\bfb$}
	\label{sec:nonlocal}

 In order to include the physical coordinates where $A$ is the magnetic moment $\mu$, the last step is to consider the case where the coordinate transformation involves derivatives of the magnetic field.  This is important because perturbative reductions, such as  those  based on Lie-transforms \cite{Cary79, Litt82,IntrGyro} or mixed variable generating functions \cite{pfirsch}, often involve derivatives to arbitrary high order in the fields. 

 So,  a more general transformation to new coordinates $(\bfq, V,\hv)\leftrightarrow(\bar\bfq, A,\hw)$ is considered:  
\bq
\bfq=\bar \bfq, \quad  \hv=\hw\,,  \quad{\rm and} \quad  V=V[A,\hw,\bfb]\,,
\label{nonlocal}
\eq
where now $V[A,\hw,\bfb]$ means a transformation that depends on $\bfb$ and, possibly, all its derivatives. Clearly, invertibility requires $V_A:=\p V/\p A\neq 0$, as in Sec.~\ref{sec:local}.  Since  the first two equations above are identities,  as before eventually $ \hv$  will be used for $\hw$ and $\bfq$ for $\bar\bfq$.

 The Jacobian for this transformation is again
\bq
dz= V^2V_A dA d\Om d^3q
=\calj dA d\Om d^3q=: \calj d\eta d^3q=:\calj dw\,.
\notag
\eq
but now   $\calj$ depends on $\bfq$ through $\bfb$ and its derivatives.

 For the chain rule on functions or functionals,  $f(\bfq, V,\hv)=\bar{f}(\bar\bfq,A,\hw)$ is varied as in Sec.~\ref{sec:local}, and all terms are the same as before, except some slight changes in the relations involving derivatives with respect to the magnetic field. Indeed, the Fr\'{e}chet derivative with respect to $\bfb$ is now a differential  operator, and care must be taken  with the  order of  terms. For instance, relation (\ref{dfdq}) becomes 
\bq
 	\frac{\p  f}{\p \bfq}
 	=\frac{\p  \bar f}{\p\bar  \bfq} 
 	-\frac{\p\bar f}{\p A} ~ \frac{1}{V_A}  ~ 
 	V_{B_i} ~ \frac{\p  B_i}{\p \bfq}~ 
 	\,,
\eq
where $V_{B_i}$ is now a differential operator acting on ${\p  B_i}/{\p \bfq}$. Formulae (\ref{nabds})-(\ref{locinner}) must be changed accordingly.

 As for relations (\ref{ddfa})-(\ref{chainp3}),  variation is performed  slightly differently this time as follows:
\bq
  	\de \bar f
  	=\de f 
  	+ f_V ~ V_{\bar \bfb}\cdot \de \bar \bfb
  	\,,
  	\notag
\eq
where  $V_{\bar \bfb}$ is the Fr\'{e}chet derivative operating on $\de \bar \bfb$.  Thus the chain rule for functional derivatives gives 
\bq
 	\frac{\de F}{\de \bfb}
 	=\frac{\de \bar{F}}{\de \bar{\bfb}} 
 	+   \int \!\!d\eta~    
 	V_{\bfb}^{\dagger}\left( \frac{\p f}{\p V}
 	\frac{\de \bar F}{\de \bar f} \right)
 	=\frac{\de \bar{F}}{\de \bar{\bfb}} 
 	+   \int\!\! d\eta~    
 	V_{\bfb}^{\dagger}\left(
	 \frac{{\bar{F}}_{\bar f}}{V_A} 
 	\frac{\p \bar f}{\p A}
 	 \right)
 	\,,
 	\notag
\eq
where  the adjoint $\dagger$ is done with respect to $dw$. 

 Finally, the Maxwell-Vlasov bracket (\ref{BrackMVCoordA}) in these  coordinates becomes 
\begin{align}
 	&\{F,G\}=  
 	\int\!\! d\eta d^3q ~ \calj  {f}
 	\left[\calj^{-1}{F}_{{f}},\calj^{-1}{G}_{{f}}\right]
 	\label{BrackMVCoordMu}
 	\\
  	&+    e \int\!\!  d\eta d^3q~ \calj {f} 
  	\left(  G_{\bfe}\cdot D_*  \calj^{-1}{F}_{ f}  
  	-  F_{\bfe}\cdot D_*  \calj^{-1}{G}_{ f}\right)
	\nonumber
	\\
  	&+ \int\!\!d^3q ~ \left( F_{\bfe}\cdot \nabla\times 
  	\left[  G_{\bfb} 
  	+  \int\!\! d\eta~   V_{\bfb}^{\dagger}\left( \frac{G_f}{V_A} 
 	\frac{\p   f}{\p A} \right)\right]
  	\right.
	\notag
	\\
  	&\hskip5ex\left.
  	-  G_{\bfe}\cdot \nabla\times  \left[ F_{\bfb}  
  	+ \int\!\! d\eta~     
  	V_{\bfb}^{\dagger}\left( \frac{F_f}{V_A} 
 	\frac{\p   f}{\p A}  \right)\right]
  	\right)
  	\,.
  	\notag
\end{align}

%%%%%%%%%%%%%%%%%%%%%%%%%%%%%%%
	\section{An example using  the magnetic moment} 
	\label{sec:example}

 With the transformed  bracket (\ref{BrackMVCoordMu}), the first thing to be checked is whether the  dynamics agrees with the conservation of the magnetic moment, when appropriate, since this is what suggested the reduction in the first place.  To this end, suppose the coordinate $A$ is the magnetic moment, $A:=\mu(\bfq,\bfv)$, which to lowest order is given by $A=\|\bfv_{\perp}\|^2/\|\bfb\|$.  To get a true conserved quantity, small corrections must be  added to all orders in the Larmor radius, including derivatives of all orders in the magnetic field \cite{CaryBriz09, IntrGyro}. Thus, $A=\|\bfv_{\perp}\|^2/\|\bfb\|+O(\epsilon)$ is defined as solution of the following equation
\[
	0=\dot\mu
	= \bfv\cdot\na \mu +e \bfv\x \bfb\cdot\p_\bfv \mu 
	\,.
\] 

 At the field level, the conservation of the magnetic moment corresponds to the conservation of the functional 
$$
	M:=\int\!\! dz ~ f \mu  
$$ 
for any particle distribution $f$. In the transformed  coordinates, this is  
$$
	\bar M:=\int\!\! dw ~ \calj \bar f \mu
	\,.  
$$ 

 To investigate the conservation of $\bar M$,  note that a static magnetic field corresponds to elimination of the electric field term in the Hamiltonian functional, since this eliminates the  $\na\x\bfe$ term in the Maxwell-Faraday equation. In this case 
\bqy
	\dot{\bar M} 
	&=& 
	\{\bar M,\bar{H} \}
	=
	\int\!\! d\eta d^3q ~ \calj  \bar f
 	\left[\mu,\calj^{-1}{ \bar{H} }_{\bar f} \right]
 	\notag\\
 	&=&
 	\tfrac{1}{2} \int\!\! d\eta d^3q ~ \calj  \bar{f}
 	\left( \na_*\mu\cdot D_* {V^2} 
  	+e\bfb \cdot D_*\mu\x D_*{V^2}\right)                
 	\notag\\
 	&=&
 	\int\!\! d^3v d^3q ~  {f}
 	\left( \bfv\cdot\na\mu 
  	+\bfv\x e\bfb \cdot \p _\bfv\mu \right) 
	=0\notag
 	\,,
\eqy
as was  expected. 

 Accordingly, the transformed  bracket (\ref{BrackMVCoordMu}) is expressed in coordinates adapted to the conserved magnetic moment. As is ususal in gyrokinetics, the electromagnetic field dynamics spoils the conservation of the magnetic moment.   This is why  the feed-back of the  plasma dynamics onto the electromagnetic field dynamics needs to  be restored as a perturbation, i.e., a perturbed magnetic moment must  be defined that is conserved \cite{BrizHahm07}.

 Consider now   the transformed  Maxwell-Vlasov equations of motion generated  by the bracket (\ref{BrackMVCoordMu}).  In this bracket, most of the terms are actually identical to those of  the initial bracket (\ref{BrackMV}), even though their formal expressions look  different because they are expressed in the reduced coordinates $(\bar\bfq, A,\hw)$, e.g. through formulae (\ref{newds}) and (\ref{locinner}). The only new  terms are 
\[
 	\int\!\!d^3q ~ \bar F_{\bar \bfe}\cdot \nabla\times 
  	\int\!\! d\eta~   
  	V_{\bar \bfb}^{\dagger}
  	\left( 
  	\frac{{\bar G}_f}{V_A} \frac{\p   \bar f}{\p A}	
 	\right)
 	\,,
\]
and one obtained by permuting $\bar F$ and $\bar G$ (and with a minus sign for bracket antisymmetry). 

 In the equations of motion, this new bracket term  generates an additional term in Maxwell-Ampere equation, viz. 
 \bqy
 \dot{\bar \bfe}  &=&
  	\nabla\times  \bar{H}_{\bar\bfb} 
 	- e \int\!\!  d\eta ~ \calj \bar{f} ~
   	  D_*  
   	\left(\calj^{-1}{\bar { H}  }_{\bar f}\right)
   	\nonumber\\
  	&{\ }& \hspace{2.0 cm} +  \nabla\times 
    \int\!\! d\eta~   V_{\bar\bfb}^{\dagger}
    \left( 
    \frac{ \bar{H}_f  }{V_A} 
    \frac{\p   \bar f}{\p A}
 	\right)
 	\,.
  	\label{MotionMagnet}
\eqy
At first glance this additional term looks  like a  new magnetization current. But,  one must  remember that the usual $\nabla\times  \bfb$ term has itself another additional contribution $\nabla\times {\de \bar{ H}_{kin}}/{\de \bar\bfb}$, because in the reduced variables, the plasma kinetic energy depends on the magnetic field $\bar {H}_{kin}:=\int\!\! dw \calj \bar f {V^2}/{2}$  that is not constant in $\bar \bfb$ (both because of $\calj$ and $V$). And,  it turns out that this last additional contribution exactly cancels the ``magnetization" term in (\ref{MotionMagnet}):
\bqy
 	\frac{\de \bar {H}_{kin}}{\de \bar\bfb(\bfx)}
 	&=&
 	\tfrac1{2}\int\!\! dw ~ \bar f 
 	\left( {\calj V^2}
 	\right)_{\bar \bfb} \de(\bfq-\bfx)
 	\notag
 	\\
 	&=&
 	\tfrac1{2}\int\!\! dw~ \bar f 
 	\left({\p _A V \cdot V^4} \right)_{\bar \bfb} 
 	\de(\bfq-\bfx)
 	\notag
 	\\
 	&=&
 	- \tfrac1{10}\int\!\! dw~ \p _A \bar f 
 	\left( {V^5} \right)_{\bar \bfb} 
 	\de(\bfq-\bfx)
 	\notag
 	\\
 	&=&
 	- \tfrac1{10}\int\!\! dw~ \p _A \bar f 
 	\left(  {V^5} \right)_V V_{\bar \bfb} 
 	\de(\bfq-\bfx)
 	\notag
 	\\
 	&=&
 	- \tfrac1{2}\int\!\! dw~ 
 	\p _A \bar f ~  {\calj V^2}{V_A} ~ V_{\bar \bfb} 
 	\de(\bfq-\bfx)
 	\notag
 	\\
 	&=&
 	- \int\!\! dw~ 
 	\p _A \bar f ~\tfrac{{ \bar{H}}_{\bar f}}{V_A} ~ V_{\bar \bfb} 
 	\de(\bfq-\bfx)
 	\notag
 	\\
 	&=&
 	- \int\!\! d\eta~ V_{\bar \bfb}^\dagger 
 	\left( 
 	\tfrac{\bar{H}_{\bar f} }{V_A}~ \p _A \bar f 
 	 \right)
 	\notag
 	\,. 
\eqy

 This cancellation was to be expected, since the electric field $\bfe=\bar \bfe$ is not affected by the change of velocity coordinates, and the current term has not been changed either, but only expressed in the new variables:
$$
  	- e \int\!\!  d\eta ~ \calj \bar{f} ~   
  	D_* \left( \calj^{-1} \bar{  H}_{\bar f}  \right)
  	=
  	- \frac{e}{2} \int\!\!  d^3v ~{f}  ~  \p _\bfv  
  	\left(\calj^{-1} \calj {\|\bfv\|^2}  \right)
  	=
  	- \mathbf J
  	\,.
$$

 Finally, the additional term in the transformed  bracket (\ref{BrackMVCoordMu}) generates another additional term in the  equation of motion: the dynamics of the   Vlasov phase space density $\dot f$ has an additional force term 
$$
	- \frac1{V_A} \frac{\p   \bar f}{\p A}~
   	V_{\bar \bfb} \cdot \na\x\bar \bfe
  	\,.
$$
This term is not cancelled by any other term.  It can  be rewritten as 
$$
  	-\frac{\p   f}{\p V}~
  	V_{\bfb}\cdot \na\x\bfe
  	=  
  	\frac{\p   f}{\p V}~
   	V_{\bfb} \cdot \dot \bfb
  	\,,
$$
which is exactly the expected contribution when applying the chain rule for the time derivative of the transformed fields. It comes about because the change of coordinates is time-dependent when the magnetic field is not static.

%%%%%%%%%%%%%%%%%%%%%%%%%%%%%%%
\section{Conclusion}
\label{sec:conclusion}

 In summary, in this paper techniques for  transforming  the Vlasov-Maxwell Poisson bracket to new coordinates, when  the transformation law mixes dependent and independent variables,  have been developed.    Four transformations were considered, each of which considered a new feature needed for  understanding the more general transformation of (\ref{nonlocal}).  In Sec.~\ref{sec:mag-coords} a transformation that mixed the independent velocity variable with the magnetic field was considered    and the associated function and functional chain rules were described.  In Sec.~\ref{sec:spherical}, spherical velocity coordinates were treated and here it was seen how a nontrivial  Jacobian determinant influences a transformation.  In Sec.~\ref{sec:local} a class of transformations  that  mixes the dependent and independent variables by having dependence on $\bfb$ and in addition  possesses a nontrivial Jacobian was considered.  Finally, in Sec.~\ref{sec:nonlocal},   the nonlocal transformation of  (\ref{nonlocal}) was effected,  the most general transformation of this paper that results  in the transformed noncanonical Poisson bracket of (\ref{BrackMVCoordMu}).   This final form of the Poisson bracket  was seen to contain additional terms  that  appear to be magnetization-like contributions. However, these bracket terms were shown to produce  no magnetization term in the equations of motion, since the electromagnetic fields are not affected by the change of field coordinates. Only the dynamics of the Vlasov density obtained an additional term, a term that  results from the change of field coordinates being time-dependent through $\bfb$.

 The transformations of  Secs.~\ref{sec:mag-coords}--\ref{sec:nonlocal} paved the way for the simple example of Sec.~\ref{sec:example}.    Here the dynamics was reduced by dropping the electric field energy from the Hamiltonian, resulting in  the magnetic moment being  conserved by a reduced dynamics that must have a static magnetic field.  However,  when restoring the feed-back of the plasma dynamics onto the electromagnetic field dynamics, the magnetic moment was  seen to be no longer  conserved and must be  perturbatively changed to be  conserved.

In all  the cases considered, the lifting was eased because the change of coordinates only concerned  a new  particle velocity that depends on the magnetic field,  but no change was made in the spatial coordinate.  If Eq.~(\ref{gentrans}) is generalized by adding dependence on the electric field and all its derivatives, then results similar to those presented are immediate.   However,  if the new spatial variable has velocity and field dependence, then the situation becomes considerably more complex.  Such transformations are of interest for some  oscillation-center,  guiding-center, and gyrokinetic theory development,  and the same methods of function and functional chain rule can be used, but some additional effects will show up, e.g., non-zero polarization and magnetization terms like those of \cite{morrison12}.   Details of  the magnetic moment reduction will be given in \cite{IntrGyro} and  more general lifting will be considered in a future publication.

 %%%%%%%%%%%%%%%%%%%%%%%%%%%%%%%
\section*{Acknowledgment}

\noindent 
We acknowledge financial support from the Agence Nationale de la Recherche (ANR GYPSI). This work was also supported by the European Community under the contract of Association between EURATOM, CEA, and the French Research Federation for fusion study. The views and opinions expressed herein do not necessarily reflect those of the European Commission.  PJM was supported by U.S. Department of Energy contract \# DE-FG05-80ET-53088. The authors also acknowledge fruitful discussions with Alain Brizard, and with members of the \'Equipe de Dynamique Nonlin\'eaire of the Centre de Physique Th\'eorique of Marseille.

%%%%%%%%%%%%%%%%%%%%%%%%%%%%%%%
%\section*{References}


\begin{thebibliography}{78}


\bibitem{celestial}
C. L. Siegel and J. K. Moser, 
{\it  Lectures on Celestial Mechanics} (Springer Verlag, Berlin, 1971).

\bibitem{born}
M. Born, {\it The Mechanics of the Atom}  (F. Ungar, New York, 1967).

\bibitem{kruskal} 
M. Kruskal, 
%Asymptotic Theory of Hamiltonian and other Systems with all Solutions
%Nearly Periodic
J. Math. Phys. {\bf 3}, 806 (1962).

\bibitem{llave}
R. de la Llave, {\it Introduction to KAM theory}, in Computational Physics (World
Scientic, River Edge, NJ, 1992)  pp. 73--105.

\bibitem{henrard}  J. Henrard,
{\it  The adiabatic invariant in classical dynamics},  Dynamics Reported 2, (Springer, Berlin, 1993) p.\ 
117 (new series).

\bibitem{morgreene}
P. J. Morrison and J. M. Greene,
 Phys. Rev. Lett. {\bf 45}, 790, (1980); E  \textbf{48}, 569 (1982).

\bibitem{Morr81}
P. J. Morrison, AIP Conf. Proc. {\bf 88}, 13 (1982).

\bibitem{morrison98}
P. J. Morrison, Rev. Mod. Phys. {\bf 70}, 467 (1998).


\bibitem{Morr80}
P. J. Morrison, Phys. Lett. {\bf 80A}, 383 (1980).

\bibitem{Wein81}
A. Weinstein and  P. J. Morrison, Phys. Lett. A {\bf 86}, 235 (1981).

\bibitem{Mars82}
J. E. Marsden and A. Weinstein, Physica D {\bf 4}, 394 (1982).

\bibitem{morrison12}  P. J. Morrison, ``A general theory of gauge-free lifting,'' arXiv:1210.6564 [physics.plasm-ph],  submitted to Phys.\  Plasmas  (2012).

\bibitem{Frieman} E. A. Frieman and L. Chen, 
Phys. Fluids {\bf 25},  502 (1982).

\bibitem{littlejohn84}
R. G. Littlejohn, 
Phys. Fluids {\bf 27},  976 (1984).


\bibitem{mischenko}
A. J. Brizard and A. Mischenko, 
J. Plas. Phys. {\bf 75},  675 (2009).

 
\bibitem{CaryBriz09}
J. R. Cary and A. J. Brizard,
Rev. Mod. Phys. {\bf 81}, 693 (2009).

\bibitem{BrizHahm07}
A. J. Brizard and T. S. Hahm, 
Rev. Mod. Phys. {\bf 79}, 421 (2007).

\bibitem{Morr05}
P. J. Morrison, Phys. Plasmas {\bf 12}, 058102 (2005).

\bibitem{IntrGyro}
L. de Guillebon, N. Tronko, M. Vittot, and Ph. Ghendrih,
``Dynamical reduction for charged particles in a strong magnetic field without guiding-center", in preparation.


\bibitem{Cary79}
J. R. Cary,
Phys. Reports {\bf 79}, 129  (1981).

\bibitem{Litt82}
R. G. Littlejohn,
J. Math. Phys. {\bf 23}, 742  (1982).

\bibitem{pfirsch}
D. Correa-Restrepo and D. Pfirsch, 
J. Plasma Phys.  {\bf 71}, 1 (2005).  
 

%%%%%%%%%%%%%%%%%%%%%%%%


% 
%
%
%\bibitem{Litt88}
%R. G. Littlejohn, 
%Phys. Rev. A {\bf 38} (1988), 6034.
%
%\bibitem{Sugi08}
%L. E. Sugiyama,
%Phys. Plasmas {\bf 15} (2008), 092112, 1.
%
%\bibitem{Nort63}
%T. G. Northrop, 
%Wiley, New York, (1963).
%
%\bibitem{Litt81}
%R. G. Littlejohn,
%Phys. Fluids {\bf 24} (1981), 1730--1749.
%
%\bibitem{SpenKauf82}
%R.G. Spencer, A.N. Kaufman,
%Phys. Rev. A {\bf 25} (1982), 2437.
%
%
% 
%
%%%%%%%%%%%
%
% 
%
%\bibitem{HolmStab85}
%D. Holm, J. Marsden, T. Ratiu and A. Weinstein, 
%Phys. Rep. {\bf 123} (1985), pp. 1.
%
%\bibitem{Scott10}
%B. Scott and J. Smirnov, 
%Phys. Plasmas {\bf 17}, 112302 (2010).
%
%\bibitem{Tronk11}
%A.J. Brizard, N. Tronko
%Phys. Plasm. {\bf 18} (2011), 082307.
%
%   
%
%\bibitem{Litt81}
%R. G. Littlejohn,
%Phys. Fluids {\bf 24} (1981), 1730--1749.
%
% 


\end{thebibliography}
\end{document}